\documentclass[12pt]{article}

\usepackage{latexsym}

\newtheorem{thm}{Theorem}

\begin{document}

\title{Exact static solutions in Einstein-Maxwell-Dilaton gravity
with arbitrary dilaton coupling parameter\thanks{A short version
of the talk given at the $4^{th}$ General Conference of the Balkan
Physical Union, Veliko Turnovo , Bulgaria, $22$-$25$ August,
$2000$.  }}

\author{Stoytcho S. Yazadjiev \thanks{E-mail: yazad@phys.uni-sofia.bg}\\
{\footnotesize  Department of Theoretical Physics,
                Faculty of Physics, Sofia University,}\\
{\footnotesize  5 James Bourchier Boulevard, Sofia~1164, Bulgaria
}\\ }

\date{}

 \maketitle

\begin{abstract}
We present  solution generating methods which allow us to
construct exact static solutions to the equations of
four-dimensional Einstein-Maxwell-Dilaton gravity  starting with
arbitrary static solutions to the pure vacuum Einstein equations,
Einstein-dilaton or Einstein-Maxwell equations.

\end{abstract}

\vskip2pc

\newcommand{\lfrac}[2]{{#1}/{#2}}
\newcommand{\sfrac}[2]{{\small \hbox{${\frac {#1} {#2}}$}}}
\newcommand{\ben}{\begin{eqnarray}}
\newcommand{\een}{\end{eqnarray}}
\newcommand{\la}{\label}


In four dimensions the field equations of the
Einstein-Maxwell-dilaton gravity with arbitrary dilaton coupling
parameter $\alpha$, can be obtained from the following Einstein
frame action \cite{GHS}

\ben
 {\cal A}= -{1\over 16\pi} \int d^4x \sqrt{-g} \left(R -
2\partial_{\mu}\varphi\partial^{\mu}\varphi   - e^{-2\alpha
\varphi}F_{\mu\nu}F^{\mu\nu}\right) \la{AMDA} \,. \een

Here $R$ is the Ricci scalar with respect to the space-time metric
$g_{\mu\nu}$ (with a signature $(-,+,+,+)$), $\varphi$ is the
dilaton field and $F_{\mu\nu}=(dA)_{\mu\nu}$ is the Maxwell
two-form.

The aim of the present work is to present  methods for generating
exact static solutions to the EMd-gravity equations generalizing
the  previous work of the author \cite{Yazadjiev}.

For a static space-time the metric can be written in the form

\ben ds^2 = - e^{2u}dt^2 + e^{-2u}h_{ij}dx^{i}dx^{j} \een

where $e^{2u}= - g(\xi,\xi)$.

In order to simplify calculations we will consider the pure
electric case. In this case the Maxwell two-form is:

\ben F= e^{-2u}\xi \wedge d\Phi \een

where $\Phi$ is the electric potential.

In terms of the three-dimensional metric $h_{ij}$ the field
equations following from the action (\ref{AMDA}) are:

\ben  \la{TFE}
 {{}^{3}R}_{ij} = 2D_{i}uD_{j}u +
2D_{i}\varphi D_{j}\varphi  - 2e^{-2u -2\alpha \varphi}D_{i}\Phi
D_{j}\Phi
\\ \nonumber
D_{i}D^{i}u = e^{-2u -2\alpha \varphi}D_{i}\Phi D^{i}\Phi
\\ \nonumber
D_{i}D^{i}\varphi = \alpha e^{-2u -2\alpha \varphi}D_{i}\Phi
D^{i}\Phi
\\ \nonumber D_{i}\left(e^{-2u -2\alpha \varphi}D^{i}\Phi\right) = 0
\, .\een

Here $D_{i}$ is Levi-Civita connection and ${{}^{3}R}_{ij}$ is the
Ricci tensor with respect to the three-metric $h_{ij}.$

Let us introduce the following symmetric matrix

\ben {\stackrel{\alpha}S} = \pmatrix{e^{(1 + \alpha)u  + (\alpha -
1)\varphi} -(1 + \alpha^2 )\Phi^2 e^{(\alpha - 1)u - (1 +
\alpha)\varphi} & -\,\sqrt{1 + \alpha^2 }\Phi e^{(\alpha - 1)u -
(1 + \alpha)\varphi}   \cr   -\,\sqrt{1 + \alpha^2 }\Phi
e^{(\alpha - 1)u - (1 + \alpha)\varphi}    & - \, e^{(\alpha - 1)u
- (1 + \alpha)\varphi} } \la{S} .\een

In the terms of the matrix ${\stackrel{\alpha}S}$ the equations
(\ref{TFE}) can be written in the following compact form:

\ben \la{MTFE}
 {{}^{3}R}_{ij} = - \,\,{1\over 1 +
\alpha^2 }
Sp\left(\partial_{i}{\stackrel{\alpha}S}\partial_{j}{\stackrel{\alpha}S}^{-1}\right)
  \nonumber \\
  D^{i}\left({\stackrel{\alpha}S}^{-1}D_{i}{\stackrel{\alpha}S}
  \right)= 0 \, .
\een

The above equations can be derived from the following action

\ben \la{MA}{\tilde {\cal A}}= \int \sqrt{h}\left({}^{3}R -
{h^{ij}\over 1+
\alpha^2}Sp\left(D_{i}{\stackrel{\alpha}S}D_{j}{\stackrel{\alpha}S}^{-1}
\right) \right)d^3x \, .\een

The action (\ref{MA}) is invariant under the group $GL(2,R)$ for
fixed projection metric $h_{ij}$. The group $GL(2,R)$ acts
explicitly as follows $$ {\stackrel{\alpha}S} \to
G{\stackrel{\alpha}S}G^{T}$$

where $G\in GL(2,R)$.

The $GL(2,R)$ symmetry can be employed for generating new exact
EMD solutions from known ones. In particular, it will be useful to
employ the $GL(2,R)$ symmetry to generate exact solutions with
nontrivial electric field from any given solution of the pure
vacuum Einstein equations or Einstein-dilaton equations (for
details see \cite{Yazadjiev1}). The action of the symmetry group
does not, in general, preserve the asymptotic flatness. The
subgroup preserving the asymptotic flatness  is $SO(1,1)$ (see
\cite{Yazadjiev2}).

Another method which allows us to generate large classes of exact
EMD solutions from solutions of pure vacuum Einstein equations
will be discussed below.

We assume now that the matrix ${\stackrel{\alpha}S}$ depends on
the space coordinates only through one potential $\chi$ satisfying
the equation

\ben D_{i}D^{i}\chi = 0 \, . \een

Then requiring also that

\ben \la{CON} -{1 \over 2\left(1 + \alpha^2 \right)}
Sp\left({d{\stackrel{\alpha}S}\over d\chi}
{d{\stackrel{\alpha}S}^{-1}\over d\chi}\right) = 1  \een

the equations (\ref{MTFE}) are reduced to the following

\ben \la{RFE}
 {{}^{3}R}_{ij} = 2 D_{i}\chi D_{j}\chi \nonumber \\
D_{i}D^{i}\chi = 0 \nonumber \\ {d \over d\chi
}\left({\stackrel{\alpha}S}^{-1}{d \over
d\chi}{\stackrel{\alpha}S}
  \right)= 0  \, .
\een

The important observation is that the first two equations of
(\ref{RFE}) are actually the static vacuum Einstein equations. The
third equation is separated from the first two and can be formally
integrated. Its general asymptotically flat solution is

\ben {\stackrel{\alpha}S}= \sigma_{3}\, e^{Q \,\chi}\,  \een

where  $Q= \pmatrix{ a & b \cr -b & c}$  and  $\sigma_{3}$ is the
third Pauli matrix. In dependence of the determinant of the matrix
${\stackrel{\alpha}S}$ we obtain three classes of solutions.

 The first class solutions is obtained when $\det Q<1 + \alpha^2
 $:
\\
 \ben e^{2u} = (1 - \Gamma^2)^{2\over 1 + \alpha^2} {e^{2\chi
\cos(\omega - \omega_{\alpha})}   \over \left(1 - \Gamma^2
e^{2\chi \sqrt{1 + \alpha^2} \cos(\omega)} \right)^{2\over 1 +
\alpha^2}} \nonumber     \; , \een \ben e^{2\varphi} = (1 -
\Gamma^2)^{2\alpha\over 1 + \alpha^2} {e^{2\chi \sin(\omega -
\omega_{\alpha})}   \over \left(1 - \Gamma^2 e^{2\chi \sqrt{1 +
\alpha^2} \cos(\omega)} \right)^{2\alpha\over 1 + \alpha^2}}   \;
, \la{ALPHAFCS} \nonumber \een \ben \Phi = {\Gamma \over \sqrt{1 +
\alpha^2}} {1 - e^{2\chi \sqrt{1 + \alpha^2} \cos(\omega)} \over 1
- \Gamma^2 e^{2\chi \sqrt{1 + \alpha^2} \cos(\omega)} } \nonumber
\; , \een където \ben \omega_{\alpha} = \arcsin\left({\alpha \over
\sqrt{1 + \alpha^2}}\right) \nonumber \, .\een

 The second class solutions is obtained for $\det Q=1 + \alpha^2$:

\ben e^{2u} = {e^{{2\alpha \over \sqrt{1 + \alpha^2}}\chi} \over
(1 - b\chi)^{{2\over 1 + \alpha^2}} }  \nonumber  \; , \een \ben
e^{2\varphi} = {e^{-{2\over \sqrt{1 + \alpha^2}}\chi} \over (1 -
b\chi)^{{2\alpha\over 1 + \alpha^2}} } \; , \la{ALPHATCS}
\nonumber \een \ben \Phi = -{1\over \sqrt{1 + \alpha^2}} {b\chi
\over 1 - b\chi} \nonumber    \; . \een

The third class solutions is obtained when $\det Q>1 + \alpha^2$:

\ben e^{2u}= e^{{2\alpha \chi\over \sqrt{1 + \alpha^2}}
\cosh(\psi)}\! {\cos^{2\over 1 + \alpha^2}(\vartheta)  \over
\cos^{2\over 1 + \alpha^2}(\chi \sqrt{1 + \alpha^2} \sinh(\psi) +
\vartheta)} \nonumber   \een \ben e^{2\varphi}=\! e^{-{2\chi \over
\sqrt{1 + \alpha^2}} \cosh(\psi)}\!\!\! {\cos^{2\alpha\over 1 +
\alpha^2}(\vartheta)  \over \cos^{2\alpha\over 1 + \alpha^2}(\chi
\sqrt{1 + \alpha^2} \sinh(\psi) + \vartheta)} \la{ALPHASCS}
\nonumber \een \ben \Phi = - {1 \over \sqrt{1 + \alpha^2}}
{\sin(\chi \sqrt{1 + \alpha^2} \sinh(\psi))  \over \cos(\chi
\sqrt{1 + \alpha^2} \sinh(\psi) + \vartheta)} \nonumber \; . \een

Here the free parameters $\Gamma^2<1$, $\omega$, $\Psi$ and
$\vartheta$ are functions of the original parameters $a$, $b$ and
$c$.

The obtained results can be summarized in the following

\begin{thm}\nonumber
Let ${g^{E}}$ is a static, asymptotically flat solution to the
Einstein equations and $e^{2\chi}= -g^{E}(\xi,\xi)$. Then the
metric  $$g= e^{2\chi-2u}{g^{E}} - \left(e^{-2u}- e^{2u - 4\chi}
\right)\xi\otimes \xi $$  together with the two form $F =
e^{-2u}\xi\wedge d\Phi$ and scalar field $\varphi$ is a static and
asymptotically flat solution to the EMD-gravity equations when
${\bf u}$, $\bf \varphi$ and $\Phi$ are given by one the classes
(1), (2) and (3).
\end{thm}

For spherically symmetric space-times most of the obtained
solutions describe globally naked strong curvature singularities
\cite{Yazadjiev}. Only within the first class solutions, for the
particular case $\omega=\omega_{\alpha}$, we obtain the charged
dilaton black hole solution.

It is also interesting  to consider static, axi-symmetric
space-times. In this case the metric can be written in the form

\ben ds^2 = - e^{2u}dt^2 + e^{2h-2u}(d\rho^2 + dz^2) +
e^{-2u}\rho^2d\phi^2  \, .\een

For static, axi-symmetric space-times the EMD equations take the
form

\ben \la{SASEMD}
\partial_{\rho}\left(\rho {\stackrel{\alpha}S}^{-1}\partial_{\rho}{\stackrel{\alpha}S}\right)
+ \partial_{z}\left(\rho
{\stackrel{\alpha}S}^{-1}\partial_{z}{\stackrel{\alpha}S} \right)
= 0  \nonumber \\ {1\over \rho} \partial_{\rho}h = {1\over 2(1+
\alpha^2 )}
\left(Sp\left(\partial_{z}{\stackrel{\alpha}S}^{-1}\partial_{z}{\stackrel{\alpha}S}
\right)   -
Sp\left(\partial_{\rho}{\stackrel{\alpha}S}^{-1}\partial_{\rho}{\stackrel{\alpha}S}
\right)\right)\\ {1\over \rho}\partial_{z}h  = - {1\over 1 +
\alpha^2} Sp \left(
\partial_{\rho}{\stackrel{\alpha}S}\partial_{z}{\stackrel{\alpha}S}^{-1}\right)
\nonumber \, .\een

The first equation of the above system is well-known. This
equation can be solved by use of the inverse scattering problem
method. This method allows us to construct $n$-soliton solutions.
The inverse scattering problem method, however, requires tedious
calculations. It seems that the method described above is more
powerful than the inverse scattering problem method  and it gives
larger classes of solutions in more direct and simpler manner.
Assuming again that the matrix ${\stackrel{\alpha}S}$ depends on
the space-coordinates only through the one harmonic potential
$\chi$ we obtain

\ben {\stackrel{\alpha}S}= \sigma_{3}\, e^{Q \,\chi}\,\nonumber
\een

where

\ben
\partial^2_{z}\chi +
{1\over \rho}\partial_{\rho}\chi + \partial^2_{\rho}\chi = 0
\nonumber \, .\een

Note that here the elements of the matrix $Q$ are not constrained
by (\ref{CON}) and  can be arbitrary.

We would also like  to discuss briefly once more method for
constructing exact EMD solutions.  One may wonder whether exact
EMD solutions can be constructing from  solutions of
Einstein-Maxwell equations. We shall demonstrate that this is
possible \cite{Yazadjiev1}, \cite{Yazadjiev2}.

Let us go back to the static, axisymmetric EMD equations
(\ref{SASEMD}) and to introduce the new potentials $U = u +
\alpha\varphi$, $\Psi=\alpha u - \varphi$, $\Lambda= \sqrt{1+
\alpha^2} \Phi$ and the new metric function $H = (1+\alpha^2) h$.
The static, axisymmetric EMD equations can be rewritten in the
form

\ben
\partial^2_{\rho}\Psi + {1\over \rho}\partial_{\rho}\Psi +
\partial^2_{z}\Psi = 0 \nonumber \\
\partial^2_{\rho}U + {1\over \rho}\partial_{\rho}U +
\partial^2_{z}U = e^{-2U}\left((\partial_{\rho}\Lambda)^2 + (\partial_{z}\Lambda)^2 \right)
\nonumber \\
\partial_{\rho}\left(\rho e^{-2U}\partial_{\rho}\Lambda \right) +
 \partial_{z}\left(\rho e^{-2U}\partial_{z}\Lambda \right) =0
\\ {1\over \rho} \partial_{\rho}H = (\partial_{\rho}U)^2 -
(\partial_{z}U)^2 + (\partial_{\rho}\Psi)^2 - (\partial_{z}\Psi)^2
- e^{-2U}\left((\partial_{\rho}\Lambda)^2 -
(\partial_{z}\Lambda)^2 \right) \nonumber \\ {1\over
\rho}\partial_{z}H = 2\partial_{\rho}U \partial_{z}U   +
2\partial_{\rho}\Psi
\partial_{z}\Psi  - 2e^{-2U}\partial_{\rho}\Lambda \partial_{z}\Lambda
\nonumber \, .\een

It is not difficult to recognize that the above system is just the
static, axisymmetric   Einstein-Maxwell equations along together
with minimally coupled scalar field $\Psi$. Therefore we obtain

\begin{thm}\nonumber
Let $U$, $\Psi$, $H$ and $\Lambda$ form a solution of  the static,
axisymmetric Einstein-Maxwell equations with minimally coupled
scalar field. Then $u=(1+ \alpha^2)^{-1}(U + \alpha\Psi)$,
$\varphi=(1 + \alpha^2)^{-1}(\alpha U - \Psi)$, $h= (1 +
\alpha^2)^{-1}H$ and $\Phi= (1 + \alpha^2)^{-1/2}\Lambda$ form a
solution of static, axisymmetric EMD equations.
\end{thm}

The line of the present investigation has been continued for the
EMD cosmological space-times in \cite{Yazadjiev1}.

Another approach to finding exact EMD solutions can be found in
\cite{MNQ}, \cite{MNER}, \cite{MNR}.

\vskip 0.5cm

 {\em \bf Acknowledgments:}
 \,\,\, The author is grateful to the organizers of the $4^{th}$
General Conference of the Balkan Physical Union  and especially to
Prof. M. Mateev for the opportunity to give this talk and for
financial support.

\end{document}